\documentclass[11pt]{article}
\usepackage{amsmath,amssymb,color,graphics,epsfig,cite}
\usepackage{amsfonts}
\newcommand{\hoch}[1]{$\, ^{#1}$}

\makeatletter
\@addtoreset{equation}{section}
\makeatother

\newcommand{\be}{\begin{equation}}
\newcommand{\ee}{\end{equation}}
\newcommand{\bea}{\setlength\arraycolsep{2pt} \begin{eqnarray}}
\newcommand{\eea}{\end{eqnarray}}
\newcommand{\nn}{\nonumber}

\def\ft#1#2{{\textstyle{\frac{\scriptstyle #1}{\scriptstyle #2} } }}
\def\fft#1#2{{\frac{#1}{#2}}}

\def\0{{\sst{(0)}}}
\def\1{{\sst{(1)}}}
\def\2{{\sst{(2)}}}
\def\3{{\sst{(3)}}}
\def\4{{\sst{(4)}}}
\def\5{{\sst{(5)}}}
\def\6{{\sst{(6)}}}
\def\7{{\sst{(7)}}}
\def\8{{\sst{(8)}}}
\def\sst#1{{\scriptscriptstyle #1}}

\begin{document}

\begin{flushright}
\hfill{}

\end{flushright}

%\vspace{25pt}
\begin{center}
{\Large {\bf G\"odel Universe from String Theory}}

\vspace{20pt}
{\bf Shou-Long Li\hoch{1}, Xing-Hui Feng\hoch{2, \ast}, Hao Wei\hoch{1} and  H. L\"u\hoch{2}}

\vspace{10pt}

\hoch{1}{\it School of Physics, \\
Beijing Institute of Technology, Beijing 100081, China}

\vspace{10pt}

\hoch{2}{\it Center for Advanced Quantum Studies, Department of Physics, \\
Beijing Normal University, Beijing 100875, China}

\vspace{40pt}

\underline{ABSTRACT}

\end{center}

G\"odel universe is a direct product of a line and a three-dimensional spacetime we call G$_\alpha$. In this paper, we show that the G\"odel metrics can arise as exact solutions in Einstein-Maxwell-Axion, Einstein-Proca-Axion, or Freedman-Schwarz gauged supergravity theories. The last allows us to embed G\"odel universe in string theory. The ten-dimensional spacetime is a direct product of a line and the nine-dimensional one of an $S^3\times S^3$ bundle over G$_\alpha$, and it can be interpreted as some decoupling limit of the rotating D1/D5/D5 intersection. For some appropriate parameter choice, the nine-dimensional metric becomes an AdS$_3\times S^3$ bundle over squashed 3-sphere. We also study the properties of the G\"odel black holes that are constructed from the double Wick rotations of the G\"odel metrics.

\vfill  {\footnotesize sllee\_phys@bit.edu.cn \ \ \ xhfengp@mail.bnu.edu.cn\hoch{\ast}\ \ \ haowei@bit.edu.cn \ \ \  mrhonglu@gmail.com}
% \email{sllee\_phys@bit.edu.cn}

\thispagestyle{empty}

\pagebreak

\tableofcontents
\addtocontents{toc}{\protect\setcounter{tocdepth}{2}}

%%%%%%%%%%%%%%%%%%%%%%%%%%%%%%%%%%%%%%%%

%\newpage
%%%%%%%%%%%%%%%%%%%%%%%%%%%%%%%%%%%%%%%%

\section{Introduction}

Closed time-like curves (CTCs) in solutions of General Relativity can pose challenges to Hawking's chronology protection conjecture. It is well known that Kerr metric has CTCs; however, they are hidden inside the event horizon, and hence obey the chronological censorship. The G\"odel metric \cite{Godel:1949ga}, a solution of cosmological Einstein gravity coupled to some uniform pressureless matter, describes a homogeneous rotating universe that has naked CTCs and has no globally spatial-like Cauchy surface.  However, the fine-tuning balance between the cosmological constant and the matter density in G\"odel universe makes it an unrealistic model to challenge the chronology protection.  Nevertheless it is a good toy model to study the effects of naked CTCs in both classical and quantum gravities.

The study of CTCs has enjoyed considerable attention with the development of the string
theory and the AdS/CFT correspondence.  Large number of supersymmetric or non-supersymmetric
G\"odel-like solutions, including also black holes and time machines, with naked CTCs were constructed in gauged supergravities in higher dimensions,
see e.g.~\cite{Barrow:1998wa,Gauntlett:2002nw,Herdeiro:2002ft, Boyda:2002ba,%
Harmark:2003ud,Takayanagi:2003ps,Brecher:2003rv,Brace:2003st,Gimon:2003ms,Israel:2003cx,
Cvetic:2005zi,Banados:2005da,Wu:2007gg,Barnich:2005kq,Peng:2009ty,Wu:2011im}.  These works raise important issues whether the problems of CTCs may be resolved by stringy or quantum considerations, or whether naked CTCs in the bulk implies the breakdown of unitarity of the field theory.

In this paper, we focus on the original four-dimensional G\"odel universe that is a direct product of a line and a three-dimensional metric \cite{Godel:1949ga}
\be
ds^2 = \ell^2 \Big(-(dt + r d\phi)^2 + \ft12 r^2 d\phi^2 + \fft{dr^2}{r^2} + dz^2\Big)\,.\label{godelmetric}
\ee
This metric is an exact solution of the Einstein equation involving some uniform pressureless matter
\be
R_{\mu\nu}-\ft12 R g_{\mu\nu}+ \Lambda g_{\mu\nu} = T^{\rm mat}_{\mu\nu}\,,\qquad
\Lambda=-\fft1{2\ell^2}\,,\qquad T^{\rm mat}_{\mu\nu}=u_\mu u_\nu\,,\label{pfluid}
\ee
where $u^\mu =(\ell^{-2},0,0,0)$. G\"odel universe has enjoyed continuous interests in
the general-relativity community, (see, e.g.~\cite{Banerjee,Bampi,Raychaudhuri:1980fd,%
Reboucas:1982hn,Reboucas:1986tz, Reboucas:2009yw,Agudelo:2016pic},) and G\"odel metric (\ref{godelmetric}) were generalized to G\"odel-type metrics where coefficient $\ft12$ in (\ref{godelmetric}) is replaced by a generic constant $\alpha$.  To avoid pedantry, we shall refer {\cal all} these metrics as G\"odel metrics, which are distinct from those afore-mentioned G\"odel-like metrics. It turns out that most theories associated with these general G\"odel metrics involve energy-momentum tensor of some uniform pressureless matter. Electric and magnetic fields periodic in $z$ as a replacement for such matter were used to construct G\"odel metric in \cite{Reboucas:1982hn}, where $z$ is a circular coordinate.  It is worth mentioning that the three-dimensional metric with $dz^2$ removed can arise as an exact solution of Einstein-Maxwell theory with a topological term $A\wedge F$ \cite{Banados:2005da}.  The embedding of the three-dimensional metric in heterotic string theory was first given in \cite{Israel:2003cx}.

The main purpose of this paper is to construct more Lagrangians that admit G\"odel metrics as exact solutions. We present three Lagrangians that admit the G\"odel metrics as solutions, all involving only the fundamental matter fields.  These are Einstein-Maxwell-Axion (EMA), Einstein-Proca-Axion (EPA) and Freedman-Schwarz \cite{Freedman:1978ra} $SU(2)\times SU(2)$ gauged supergravity theories. The Freedman-Schwarz model can be obtained from the effective actions of string theories via the Kaluza-Klein reduction on $S^3\times S^3$.  Thus the four-dimensional G\"odel universe can be embedded in string theory.

The paper is organized as follows. In section 2, we give a review of G\"odel metrics and their properties.  We also show that for appropriate choice of parameter, such a metric can also describe a direct product of time and a squashed 3-sphere.  In section 3, we construct EMA and EPA theories that admit G\"odel metrics, and also solutions involving squashed 3-sphere.  We generalize the theories to higher dimensions and also give the effective three-dimensional theories by Kaluza-Klein reduction.  In section 4, we consider double Wick rotations and obtain two types of black hole solutions.  In section 5, we show that G\"odel metrics are exact solutions of the Freedman-Schwarz model and hence obtain the corresponding ten-dimensional solutions of string theories.  We conclude the paper in section 6.

\section{G\"odel universe}

\subsection{The metrics of G$_\alpha \times \mathbb R$}

In this paper, we consider a class of metrics in the following form
\be
ds^2 = \ell^2 \Big(-(dt + r d\phi)^2 + \alpha r^2 d\phi^2 + \fft{dr^2}{r^2} + dz^2\Big)\,,\label{gvac}
\ee
where $\ell$ and $\alpha$ are constants.
The metric is a direct product of $\mathbb{R}$ associated with the coordinate $z$ and the three-dimensional metric of $(t,\phi,r)$, which we shall call G$_\alpha$. The original G\"odel metric \cite{Godel:1949ga} is recovered when we take $\alpha=\ft12$, corresponding to G$_{1/2}\times \mathbb{R}$.  To avoid pedantry, we shall refer the metric (\ref{gvac}) with generic $\alpha$ also as the G\"odel metric.

In addition to the constant shifting symmetry along the $(t,\phi,z)$ directions,
the G\"odel metric (\ref{gvac}) is invariant under the constant scaling
\be
r\rightarrow \lambda r\,,\qquad \phi\rightarrow \lambda^{-1} \phi\,.\label{scaling}
\ee
Note that imposing this scaling symmetry also implies that $\phi$ describes a real line, rather than a circle.  This scaling property indicates that the metric is homogeneous.
Since the metric G$_\alpha$ is three dimensional, its curvature is completely determined by the Ricci tensor, whose non-vanishing components are given by
\be
R^{\bar 0\bar 0} =\fft{1}{2\alpha \ell^2}\,,\qquad R^{\bar 1\bar 1}=\fft{1-2\alpha}{2\alpha \ell^2}=R^{\bar 2\bar 2}\,.
\label{ricci-curv}
\ee
Here we present the curvature in tangent space with the vielbein
\be
e^{\bar 0} = \ell (dt + r d\phi)\,,\qquad e^{\bar 1}=\ell \sqrt{\alpha}\,r d\phi\,,\qquad
e^{\bar 2}= \fft{\ell dr}r\,,\qquad e^{\bar 3}=\ell dz\,.\label{viel}
\ee
It is clear that when $\alpha=1$, the metric is locally AdS$_3$ (3-dimensional anti-de Sitter spacetime), i.e. G$_1=$ AdS$_3$.

\subsection{Energy condition and $\alpha$ value}

It is convenient to define the energy-momentum tensor in the vielbein basis (\ref{viel})
\be
T^{ab} = R^{ab} - \ft12 \eta^{ab} R\,.
\ee
We find $T^{ab}={\rm diag}\{\rho,p,p,\tilde p\}$, with
\be
\rho=\fft{3 -4\alpha}{4\alpha \ell^2}\,,\qquad p= \fft{1}{4\alpha\ell^2}\,,\qquad
\tilde p = \fft{4\alpha-1}{4\alpha \ell^2}\,.
\ee
The original $\alpha=\ft12$ case gives rise to matter with uniform pressure \cite{Godel:1949ga}.  The null-energy condition requires that
\be
\rho + p = \fft{1-\alpha}{\alpha \ell^2}\ge 0\,,\qquad \rho + \tilde p = \fft{1}{2\alpha \ell^2}\ge 0\,,
\ee
which is satisfied by $0< \alpha\le 1$.  As we shall see presently this implies that G\"odel metrics in general have naked CTCs in the framework of Einstein gravity.

\subsection{Metrics asymptotic to G$_\alpha$ }

We now consider deformations of G$_\alpha$ by introducing a function of two constants
\be
f=1 + \fft{a}{r} - \fft{b}{r^2}\,,\label{f}
\ee
and deform the metric (\ref{gvac}) to become
\be
ds^2 = \ell^2 \Big(-(dt + r d\phi)^2 + \alpha r^2 f d\phi^2 + \fft{dr^2}{r^2 f} + dz^2\Big)\,.\label{deformed}
\ee
It is straightforward to verify that the curvature tensors (\ref{ricci-curv}) remain unchanged.  This implies that the metric (\ref{deformed}) is locally the same as (\ref{ricci-curv}).  However, globally, the deformed metric (\ref{deformed}) is different from (\ref{gvac}). An important difference is that in the deformed metric (\ref{deformed}), the coordinate $\phi$ is periodic, namely
\be
\Delta \phi = \fft{4\pi}{\sqrt{\alpha}\, r_0^2 f'(r_0)}\,,\qquad {\rm  with}\qquad f(r_0)=0\,.
\label{phip}
\ee
This ensures that the metric is absent from a conical singularity at $r=r_0$.  After imposing this condition, the coordinate transformation that relates (\ref{gvac}) to (\ref{deformed}) breaks the scaling symmetry (\ref{scaling}) and hence the two metrics are not equivalent globally.

  To demonstrate this explicitly, we note that the parameter $a$ is trivial in that
it can be eliminated by the coordinate transformation $r\rightarrow r-\ft12a$, without altering the global structure.  The positive parameter $b$ can be set to $1$ without loss of generality, using the scaling $r\rightarrow \sqrt{b}\,r$, together with appropriate scalings of the rest coordinates.  Now let $r=\cosh\hat r$, $\phi=\hat \phi/\sqrt\alpha$, $t=\hat t-\hat \phi/\sqrt\alpha$, the metric (\ref{deformed}) with $a=0$ and $b=1$ becomes
\be
ds^2 = \ell^2\Big(-  \big(d\hat t + \ft{2}{\sqrt{\alpha}}\,\sinh^2(\ft12 \hat r)\, d\hat \phi\big)^2 + \sinh^2\hat r\, d\hat \phi^2 + d\hat r^2 + dz^2\Big)\,.\label{godel2}
\ee
On the other hand, we find that under the coordinate transformation
\bea
&&r=\cosh \hat r + \cos\hat \phi\, \sinh \hat r\,,\qquad
r\phi =\fft1{\sqrt\alpha} \sin\hat\phi\,\sinh\hat r\,,\nn\\
&& \tan\big(\ft12\hat \phi + \ft12\sqrt\alpha\, (t - \hat t)\big) = e^{\hat r}
\tan(\ft12 \hat \phi)\,,\label{trans}
\eea
the metric (\ref{gvac}) becomes also precisely (\ref{godel2}).  (This coordinate transformation reduces to the one obtained in \cite{Godel:1949ga} for $\alpha=\ft12$.)  Thus it becomes clear that when $\Delta\hat\phi=2\pi$ is fixed, the scaling symmetry of the vacuum (\ref{gvac}) is broken by the coordinate transformation (\ref{trans}), and hence the two metrics are not globally equivalent.  An important consequence is that the metric (\ref{godel2}) has CTCs for $r> r_c$ with
\be
\tanh(\ft12 r_c) =\sqrt{\alpha}\,.\label{vls}
\ee
Here $r=r_c$ is the velocity of light surface (VLS) for which $g_{\hat\phi\hat\phi}=0$. We shall call the metric (\ref{deformed}) as the deformed G\"odel metric that is asymptotic to the G\"odel metric.  For general parameters $(a,b)$, there can be two VLS's between which $g_{\phi\phi}>0$. The global structure of the metric (\ref{deformed}), written in somewhat different parametrization, was analysed in \cite{Banados:2005da}.

The emerging of the CTCs in G\"odel metrics is a consequence of that $\alpha \le 1$. If one allows $\alpha>1$, equation (\ref{vls}) has no real solution for $r_c$, and the metrics do not have naked CTCs.  However, as we saw earlier that in the framework of Einstein gravity, $\alpha>1$ violates the null energy condition.  In higher-derivative gravity due to the $\alpha'$-correction of string theory, solutions with $\alpha>1$ were constructed in \cite{Barrow:1998wa}.  However, the theory, when treated on its own, involves inevitable ghost modes.

\subsection{Mass and angular momentum}

The general G\"odel metric has two Killing vectors
\be
\xi_t = \fft{1}{\ell^2} \fft{\partial}{\partial t}\,,\qquad \xi_\phi = \fft{1}{\ell} \fft{\partial}{\partial \phi}\,.
\ee
(The Killing symmetry in $z$ direction can be broken by the matter sector in some solutions.) Following the Wald formalism \cite{Wald:1993nt}, which computes the variation of the on-shell Hamiltonian $\delta {\cal H}$ associated with a Killing vector with respect to the integration constants of the solutions, we read off the associated conversed quantities by evaluating $\delta {\cal H}$ at asymptotic infinity, and find
\be
M= \fft{\sqrt{\alpha}\,a}{16\pi}\Delta\phi\,,\qquad
J=\fft{\ell \sqrt{\alpha}\,b}{16\pi} \Delta \phi\,,
\ee
where $\Delta\phi$ is given by (\ref{phip}).  Note that if one chooses a fixed period $\Delta\phi=2\pi$ instead, rather than given by (\ref{phip}), the solution will have naked singularity at $r=r_0$, for generic $\alpha$.  We have also set the convention $\int dz=1$.  For periodic $z$, this means period $\Delta z=1$; for real line $z$, this implies that the extensive quantities such as $M$ and $J$ are in fact uniform densities over the line $z$.

\subsection{Squashed 3-sphere $S^3_\alpha \times \mathbb T$}

In the G\"odel metrics, if we let $\ell^2 =-\tilde \ell^2<0$, the metric (\ref{deformed}) becomes
\be
ds^2 = \tilde \ell^2 \Big((dz + r d\phi)^2 + \alpha h d\phi^2 + \fft{dr^2}{h} - dt^2\Big)\,,
\ee
where we have swapped the role of $(t,z)$, and set, without loss of generality, $h=1-r^2$.  Let $r=\cos\theta$, we have
\be
ds^2 = \tilde \ell^2 \Big((dz + \cos\theta d\phi)^2 + \alpha \sin^2\theta d\phi^2 +d\theta^2 - dt^2\Big)\,.\label{squashs3}
\ee
This metric describes a direct product of time with a squashed 3-sphere, which we call $S_\alpha^3$.  When the squashing parameter $\alpha=1$, $S_\alpha^3$ becomes the round $S^3$, written as a $U(1)$ bundle over $S^2$.  The regularity of $S_\alpha^3$ requires that
\be
\Delta\phi=\fft{2\pi}{\sqrt{\alpha}}\,,\qquad \Delta z = \fft{4\pi}{\sqrt{\alpha}}\,.
\ee
The period $\Delta z$ can be divided by a natural number $n$ without introducing any singularity to the manifold, giving rise to $S_\alpha^3/\mathbb Z_n$.  (Such compact G\"odel universe was also considered in \cite{Israel:2004vv}.)

\section{G\"odel solutions from Lagrangian formalism}

As mentioned in the introduction, theories in literature associated with G\"odel universe (\ref{gvac}) or (\ref{deformed}) typically involve a matter energy-momentum tensor with unknown Lagrangian origin.  The known example in four dimensions is the Einstein-Maxwell theory with an axion and a negative cosmological constant \cite{Reboucas:1982hn}
\be
{\cal L}=\sqrt{-g} \Big( R -2\Lambda- \ft12 (\partial\chi)^2 - \ft14 F^2)\,,\label{lag0}
\ee
where $F=dA$ is the field strength. For the metric (\ref{gvac}), the solutions for matter fields are
\be
A=\sqrt{2(1-\alpha)}\,\ell\, \sin(\fft{z}{\sqrt{\alpha}})\, (dt + r d\phi)\,,\qquad \chi =\fft{z}{\sqrt{\alpha}}\,,\qquad \ell^2 = -\fft{1}{2\Lambda}\,.
\ee
In this case, the continuous shifting symmetry along the $z$-direction is broken to a discrete symmetry by the Maxwell potential $A$ that is periodic in $z$. (The axion $\chi$ field does not break this symmetry since only $d\chi$ appears in the theory.)  The solution is best described as G$_\alpha\times S^1$ rather than G$_\alpha\times \mathbb R$. A consequence is that G$_\alpha$ is not a solution to the three-dimensional massless sector in the Kaluza-Klein reduction of (\ref{lag0}) on $z$.  In this section, we shall construct more examples of Lagrangians that admit the G\"odel metrics of G$_\alpha \times \mathbb R$ as exact solutions.

\subsection{Einstein-Maxwell-Axion theory with a topological term}\label{sec3}

In addition to (\ref{lag0}), we introduce an additional topological term:
\be
{\cal L} = \sqrt{-g} \left(R- 2\Lambda -\ft14 F^2 -\ft12\left(\partial \chi\right)^2 \right) +\ft18 \varepsilon^{\mu\nu\rho\sigma}\, \chi\,F_{\mu\nu} F_{\rho\sigma} \,, \label{eq23}
\ee
where $F=dA$ is the field strength, $\varepsilon^{\mu\nu\rho\sigma}$ is the density of Levi-Civita tensor whose components are $\pm1, 0$. We choose the convention $\varepsilon^{0123}=1$. The axion, Maxwell and Einstein equations of motion are given by
\begin{eqnarray}
&&\partial_\mu \left(\sqrt{-g} \partial^\mu\chi\right) +\ft18 \varepsilon^{\mu\nu\rho\sigma} F_{\mu\nu} F_{\rho\sigma}=0 \,,\qquad \partial_\mu\left(\sqrt{-g} F^{\mu\nu} - \ft12 \chi  \varepsilon^{\mu\nu\rho\sigma} F_{\rho\sigma} \right) =0 \,,  \nn \\
&&R_{\mu\nu} -\frac{1}{2}g_{\mu\nu} R +g_{\mu\nu}\Lambda -\ft12(F_{\mu\rho}{F_{\nu}}^{\rho} -\ft14g_{\mu\nu}F^2) -\ft12(\partial_\mu\chi\partial_\mu\chi -\ft12 g_{\mu\nu}(\partial\chi)^2) =0\,. \label{eq24}
\end{eqnarray}
For the general G\"odel metric (\ref{deformed}) with (\ref{f}), we consider the following ansatz for the axion and Maxwell field
\begin{equation}
A = q r\, d\phi \,, \qquad  \chi = k z\,.\label{eq27}
\end{equation}
We find that the equations of motion are all satisfied provided that
\be
k = \frac{1}{\sqrt{\alpha}},\quad \alpha= 1-\frac{q^2}{2 \ell^2},\quad \ell^2= -\frac{1}{2\Lambda}\,.\label{eq28}
\ee
The general solution contains three integration constants, $(a,b,q)$. The reality condition requires that $|q|< \sqrt2\,\ell$. It follows that we have $0<\alpha\le 1$ and $k\ge 1$.  The original $\alpha=\ft12$ G\"odel metric corresponding to $q=\ell$.  The AdS$_3$ factor arises when $\alpha=1$, corresponding to turning off the Maxwell field.  In section 3.3, we consider the case with $\alpha <0$, for which the metric describes $S_\alpha^3\times \mathbb T$.

It is worth pointing out that in four dimensions, the axion $\chi$ is Hodge dual to a 2-form potential $B_\2$ with
\be
G_\3 =dB_\2= {*d\chi} + \ft12 A\wedge F\,.\label{dual}
\ee
The Lagrangian (\ref{eq23}) is equivalent to
\be
{\cal L} = \sqrt{-g} \Big(R - 2\Lambda - \ft14 F^2 - \ft1{12} G_\3^2\Big)\,.
\ee
For the G\"odel solutions we have $B_\2 = r dt\wedge d\phi$.  The 3-form field strength $G_\3$ is suggestive of string theory, which we shall discuss in section 5.

\subsection{Einstein-Proca-Axion theory}

In this subsection, we replace the previous Maxwell field by a Proca field of mass $\mu$, with the Lagrangian
\be
{\cal L}_2 = \sqrt{-g} \left(R- 2\Lambda -\ft14 F^2 -\ft12\mu^2 A^2 -\ft12\left(\partial \chi\right)^2 \right) \,. \label{eq31}
\ee
The equations of motion are
\bea
&&\Box \chi =0\,,\qquad \nabla_\mu F^{\mu\nu} - \mu^2 A^\nu =0\,,\cr
&&R_{\mu\nu} -\ft{1}{2}g_{\mu\nu} R +g_{\mu\nu}\Lambda -\ft{1}{2}\left(F_{\mu\rho}{F_{\nu}}^{\rho} -\ft{1}{4}g_{\mu\nu}F^2\right) -\ft12\mu^2\left(A_\mu A_\nu -\ft{1}{2}g_{\mu\nu}A^2 \right) \cr
&&\quad -\ft{1}{2}\left(\partial_\mu\chi\partial_\mu\chi -\ft{1}{2}g_{\mu\nu}(\partial\chi)^2\right) =0\,. \label{eq32}
\eea
(There should be no confusion between the Proca mass $\mu$ and the spacetime indices.) The metric ansatz is given in (\ref{deformed}).  We consider the following ansatz for $A$ and $\chi$:
\be
A= q (dt + r d\phi) \,, \qquad  \chi = z\,.
\ee
Substituting these into the equations of motion, we find
\be
\mu= \frac{1}{\ell\sqrt{\alpha}}, \quad \Lambda= \frac{3q^2-2\ell^2}{4\ell^2 (\ell^2-q^2)} , \quad \alpha = 1- \frac{q^2}{\ell^2} \,.\label{eq36}
\ee
Note that all the constants except for $(a,b)$ appearing in the solution are fixed by the coupling constants of the theory, namely $(\Lambda, \mu)$.  It follows that unlike in the earlier EMA theory, the general solution involves only two integration constants. The original $\alpha=\ft12$ G\"odel metric corresponds to taking $q=\ell/\sqrt2$.

\subsection{The embedding of squashed 3-sphere}

In the embedding of the G\"odel metric in both EMA and EPA theories discussed above, the cosmological constant $\Lambda$ is negative.  When it is positive, the metric G$_\alpha\times \mathbb R$ becomes $S_\alpha^3\times \mathbb T$, as in (\ref{squashs3}).  For the EMA theory, we have
\begin{equation}
A = q \cos\theta\, d\phi \,, \qquad  \chi = \fft{t}{\sqrt{\alpha}}\,, \qquad \tilde \ell^2= \fft{1}{2\Lambda}\,,\qquad \alpha= 1+\fft{q^2}{2 \tilde \ell^2}\,.
\end{equation}
For the EPA theory, we have
\be
A=q (dz + \cos\theta\, d\phi) \,, \quad  \chi = t\,,\quad
\mu^2= -\frac{1}{\tilde \ell^2\alpha}, \quad \Lambda= \frac{3q^2+2\tilde\ell^2}{4\tilde \ell^2 (\tilde \ell^2+q^2)} , \quad \alpha = 1+\frac{q^2}{\tilde \ell^2} \,.
\ee
Thus we see that the embedding of the $S_\alpha^3$ in EPA theory requires a tachyonic vector with $\mu^2<0$, while in the EMA theory, no exotic matter is required.

\subsection{Generalizing to higher dimensions}

In this section, we generalize the G\"odel universe to higher dimensions by considering G$_\alpha\times \mathbb{R}^n$, namely
\be
ds^2 = \ell^2 \Big(-(dt + r d\phi)^2 + \alpha r^2 f d\phi^2 + \fft{dr^2}{r^2 f} + dz^idz^i\Big)\,,\qquad i=1,2,\ldots,n.\label{deformed2}
\ee
The $\alpha=\ft12$ solution can be still solved by (\ref{pfluid}), but with $u^\mu=(\ell^{-2},0,\ldots,0)$.

In order for the metrics to be solutions of some Lagrangians, we can replace the axion $\chi$ in the previous subsections by a $(n-1)$-form potential $B_{(n-1)}$ with the field strength $G_{(n)}=dB_{(n-1)}$.  The Lagrangian (\ref{eq23}) becomes
\be
{\cal L} = \sqrt{-g} \left[R- 2\Lambda -\ft14 F^2 -\ft1{2\,n!} G_{(n-1)}^2 \right] +\ft1{8\,(n-1)!} \varepsilon^{\mu\nu\rho\sigma\alpha_1\cdots\alpha_{n-1}}\, B_{\alpha_1\cdots\alpha_{n-1}}\,F_{\mu\nu} F_{\rho\sigma} \,.
\ee
(This should not be confused with dualizing $d\chi$ to the 3-form field strength in four dimensions, discussed in the end of subsection 3.1.) The axion ansatz (\ref{eq27}) is replaced by
\be
G_{(n)}= k\, dz_1\wedge\cdots dz_n\,.\label{gn}
\ee
The Lagrangian (\ref{eq31}) is now replaced by
\be
{\cal L}_2 = \sqrt{-g} \left(R- 2\Lambda -\ft14 F^2 -\ft12\mu^2 A^2 -\ft1{2\,n!} G_{(n)}^2 \right) \,.
\ee
The corresponding ansatz for $G_{(n)}$ is given by (\ref{gn}) with $k=1$.

\subsection{Effective three-dimensional theories}

The non-trivial part of the G\"odel universe is the three-dimensional metric G$_\alpha$. For the solutions in subsections 3.1 and 3.2, the Killing symmetry in $z$ direction is maintained by the matter fields.  We can thus perform dimensional reduction on the coordinate $z$.  The EMA theory becomes
\be
{\cal L} = \sqrt{-g} \left(R- 2\Lambda_{\rm eff} -\ft14 F^2 \right) +\ft14 \lambda_{\rm eff} \,\varepsilon^{\mu\nu\rho} A_\mu F_{\nu\rho} \,,
\ee
In this case, the three-dimensional G$_\alpha$ metric is now supported by
\be
A=q r d\phi\,,\qquad \alpha = 1 - \fft{q^2}{2\ell^2}\,,\qquad \Lambda_{\rm eff} = \fft{q^2-\ell^2}{2\ell^2(2\ell^2-q^2)}\,,\qquad \lambda_{\rm eff}^2 = \fft{2}{2\ell^2 - q^2}\,.
\ee
The G$_\alpha$ metric of this theory was constructed in \cite{Banados:2005da}, where the global structure of G$_\alpha$ was discussed. Under the Kaluza-Klein reduction, the EPA theory becomes
\be
{\cal L} = \sqrt{-g} \left(R- 2\Lambda_{\rm eff} -\ft14 F^2 -\ft12\mu^2 A^2\right) \,.
\ee
In this case, the G$_\alpha$ metric is supported by
\be
A= q (dt + r d\phi)\,,\qquad \mu= \frac{1}{\ell\sqrt{\alpha}}, \qquad  \alpha = 1- \frac{q^2}{\ell^2} \,,\qquad
\Lambda_{\rm eff}= \frac{2q^2-\ell^2}{4\ell^2 (\ell^2-q^2)}\,.
\ee
In both theories, the free parameters of the solutions are $(a,b)$ of function $f$ (\ref{f}), whilst $q$ and hence $\alpha$ are fixed by the coupling constants of the theories. In the above dimensional reductions, we have performed further consistent truncations to subset of fields that are relevant to the G$_\alpha$ metrics.
Note that the solution at the beginning of this section involves the $z$-dependent $A$, and hence it cannot be reduced to the three-dimensional massless sector.

\section{Black holes from double Wick rotations}

\subsection{Type I}

As was discussed in \cite{Banados:2005da}, the metrics G$_\alpha$ can describe black holes in three dimensions after double Wick rotations
\be
t\rightarrow {\rm i} t\,,\qquad \phi\rightarrow {\rm i} \phi\,.
\ee
The general metric (\ref{deformed}) now becomes
\be
ds^2 = \ell^2 \Big((dt + r d\phi)^2 - \alpha r^2 \tilde f d\phi^2 + \fft{dr^2}{r^2 \tilde f} + dz^2\Big)\,,\label{bhmetric}
\ee
with the function $\tilde f$ now given by
\be
\tilde f=1-\fft{a}{r} - \fft{b}{r^2}\,.
\ee
For the EMA theory, we find that the matter fields are given by
\be
A = q r\, d\phi \,, \quad  \chi =\fft{z}{\sqrt{\alpha}}\,,\quad\hbox{with}\quad
\ell^2= -\frac{1}{2\Lambda},\quad \alpha= 1+\frac{q^2}{2 \ell^2}\ge 1\,.
\ee
For the EPA theory, we have
\be
A= q (d\phi + r dt) \,, \quad  \chi = z\,,\quad\hbox{with}\quad
\mu= \frac{1}{\ell\sqrt{\alpha}}, \quad \Lambda= \frac{3q^2+2\ell^2}{4\ell^2 (\ell^2+q^2)} , \quad \alpha = 1+ \frac{q^2}{\ell^2}\ge 1 \,.
\ee

Thermodynamics of three-dimensional black holes was studied in \cite{Banados:2005da}.  Here we would like to derive the first law in our context and notations.  New subtlety emerges in the EMA theory, where the parameter $q$ is an integration constant. For simplicity, we shall set $\ell=1$ for the following discussions.  We also assume that the coordinate $\phi$ is periodic with $\Delta\phi=2\pi$. Thus the solution describes a rotating metric.  The null-Killing vector on the horizon $r=r_0$ with $f(r_0)=0$ is given by
\be
\xi=\fft{\partial}{\partial t} - \Omega_+ \fft{\partial}{\partial \phi}\,,\qquad \Omega_+=\fft{1}{r_0}\,.
\ee
It is straightforward to verify that the surface gravity and hence the temperature are given by
\be
\kappa = \fft{\sqrt{\alpha} r_0}{4\pi} \tilde f'(r_0)\,,\qquad T= \fft{\kappa}{2\pi}\,.
\ee
The mass and angular momentum can be read off from the Wald formalism, given by
\be
M=\ft18 \sqrt{\alpha}\, a\,,\qquad J=\ft18 \sqrt{\alpha}\, b\,.
\ee
Two situations emerge at this stage.  For black holes of the EPA theory or the effective theories in three dimensions, the parameter $q$ and hence $\alpha$ are fixed constants.  In these cases, the first law of black hole thermodynamics reads
\be
dM=T dS + \Omega_+ dJ\,.
\ee
In the EMA theory; on the other hand, the parameter $q$ is an integration constant, and hence it can be varied and should be involved in the first law.  To complete the first law involving the parameter $q$, we first note that the electric charge of the Maxwell field vanishes, namely
\be
\int {*F} + d\chi \wedge A=0\,.
\ee
(In \cite{Banados:2005da}, an electric charge associated with pure gauge transformation of $A$ was introduced. We shall not consider this here.) The linear charge density of the axion field on the other hand is non-vanishing
\be
Q_\chi =\ft18\int d\chi = \fft1{8\sqrt{\alpha}}\,.\label{qchi}
\ee
The corresponding thermodynamical potential can be read off from the 2-form potential $B_\2$ that is Hodge dual to the axion, as in (\ref{dual}).  It is given by
\be
\Phi_\chi=r_0\,.\label{axionp}
\ee
We find the first law reads
\be
dM = T dS + \Omega_+ dJ + \Phi_\chi d(\alpha Q_\chi)\,.
\ee
It is puzzling that an extra factor $\alpha$ is needed for the completion of the first law above.

\subsection{Type II}

In this interpretation,  we switch $t$ and $\phi$ in (\ref{bhmetric}) and write the metric as
\be
ds^2 = \ell^2 \Big((d\phi + (r-r_0) dt)^2 - \alpha r^2 \tilde f dt^2 + \fft{dr^2}{r^2 \tilde f} + dz^2\Big)\,.\label{bhmetric2}
\ee
Note that we also made a coordinate transformation so that the null Killing vector at the degenerate surface $r=r_0$ with $\tilde f(r_0)=0$  is $\xi=\partial_t$.  In other words, the metric is non-rotating on the horizon. The temperature is given by
\be
T=\fft{\sqrt{\alpha} r_0^2 \tilde f'(r_0)}{4\pi}\,.
\ee
The solution has no CTCs since $g_{\phi\phi}=\ell^2$ and further more $g_{tt}>0$ for $r>r_0$, and hence $t$ is globally defined outside the horizon.  Note that in this case, the entropy is a constant since the radius of the $\phi$ circle is constant.  We can also show, using the Wald formalism that the mass and angular momentum both vanish.  The solution can be viewed as thermalized vacuum. In the EMA theory, $q$ is an integration constant, which leads to non-zero electric charge and potential, give by
\be
Q_A= \fft{1}{16\pi}\int {*F} + d\chi \wedge A=\fft{q}{16\pi\sqrt{\alpha}} \int d\phi dz = \fft{q}{16\pi\sqrt{\alpha}}\Delta\phi\,,\qquad \Phi_A=q r_0\,.
\ee
The axion charge and its thermodynamical potential are given by (\ref{qchi}) and (\ref{axionp}).  This leads to the first law of black hole ``thermodynamics''
\be
\Phi_A dQ_A + \Phi_\chi dQ_\chi=0\,,
\ee
provided that $\Delta\phi=\pi$.

\section{Embedding in string theory}

\subsection{Freedman-Schwarz model}
In section 3 we constructed some ad hoc theories that admit G\"odel metrics as exact solutions. The Maxwell and axion fields are common occurrence in supergravities, indicating that there may exist an exact embedding of G\"odel universe in supergravity and hence in string theory.  In this section, we consider Freedman-Schwarz $SU(2)\times SU(2)$ gauged supergravity whose bosonic sector consists of the metric, a dilaton $\varphi$, an axion and two $SU(2)$ Yang-Mills fields. After truncating to the $U(1)^2$ subsector, the corresponding Lagrangian is
\bea
{\cal L} &=& \sqrt{-g} \Big(R - \ft12 (\partial \varphi)^2 - \ft12 e^{2\varphi} (\partial \chi)^2 + 2 (g_1^2 + g_2^2) e^{\varphi}- \ft14 e^{-\varphi} (F_1^2 + F_2^2)\Big)\cr
&&+\ft18 \varepsilon^{\mu\nu\rho\sigma}\, \chi\,(F_{1\,\mu\nu} F_{1\,\rho\sigma} + F_{2\,\mu\nu} F_{2\,\rho\sigma})\,,
\eea
where $(g_1,g_2)$ are the gauge coupling constants of the two $SU(2)$ Yang-Mills fields.
The theory admits the general deformed G\"odel metric (\ref{deformed}) with the matter fields given by
\be
A_i=q_i r d\phi\,,\qquad \chi=\fft{z}{\sqrt{\alpha}}\,,\qquad \varphi=0\,,
\ee
with the parameters
\be
\ell^2 = \fft{1}{2(g_1^2 + g_2^2)}\,,\qquad \alpha= 1 - (g_1^2 + g_2^2) (q_1^2 + q_2^2)\,.\label{ellalpha}
\ee

\subsection{G\"odel universe from string theory}

Freedman-Schwarz model can be obtained from the Kaluza-Klein reduction on $S^3\times S^3$ \cite{Chamseddine:1997mc,Cvetic:1999au}.  The relevant part of the effective Lagrangian of strings in ten dimensions is
\be
{\cal L}_{10}=\sqrt{-g} \Big( R - \ft12 (\partial \Phi)^2 -\ft1{12} e^{-\Phi} F_\3^2\Big)\,,
\ee
where $F_\3$ can be either NS-NS or R-R fields. Following the reduction ansatz given in \cite{Cvetic:1999au}, we find that the ten-dimensional solution is given by $\Phi=0$, together with
\bea
ds_{10}^2 &=& \fft{1}{g_1^2 + g_2^2} \Big(-(dt + r d\phi)^2 + \alpha r^2 f\, d\phi^2 + \fft{dr^2}{r^2f} + dz^2\Big)\nn\\
&&+ \fft{1}{g_1^2} \Big( (d\psi_1 + \cos\theta_1\, d\phi_1 + g_1 q_1 r d\phi)^2 + d\theta_1^2 + \sin^2\theta_1\,d\phi_1^2\Big) \nn\\
&&+ \fft{1}{g_2^2} \Big( (d\psi_2 + \cos\theta_2\, d\phi_2 + g_2 q_2 r d\phi)^2 + d\theta_2^2 + \sin^2\theta_2\,d\phi_2^2\Big)\,,\\
F_\3 &=& \fft{1}{g_1^2 + g_2^2} dt\wedge dr\wedge d\phi - \fft{\sin\theta_1}{g_1^2} d \theta_1\wedge d\phi_1\wedge
(d\psi_1 + \cos\theta_1\,d\phi_1 + g_1 q_1 r d\phi)\nn\\
&& - \fft{\sin\theta_2}{g_2^2} d\theta_2\wedge d\phi_2\wedge
(d\psi_2 + \cos\theta_2\,d\phi_2 + g_2 q_2 r d\phi)\nn\\
&& -\fft{q_1}{g_1} dr\wedge d\phi \wedge (d\psi_1 + \cos\theta_1\, d\phi_1)
-\fft{q_2}{g_2} dr\wedge d\phi \wedge (d\psi_2 + \cos\theta_2\, d\phi_2)\,.\label{10dsol}
\eea
Here $\alpha$ is again given by (\ref{ellalpha}).
The solution involves both electric string and magnetic fivebrane/fivebrane charges, given by
\bea
\hbox{electric}:&& Q_1\sim \int {*F}_\3 \sim \fft{1}{\sqrt{\alpha} g_1^3 g_2^3}\,,\cr
\hbox{magnetic}:&& Q_{5/5}\sim \int F_\3 \sim \fft{1}{g_1^2} + \fft1{g_2^2}\,.
\eea
These can be either all NS-NS charges or R-R charges, and the latter corresponds to the D1/D5/D5 configuration. When $q_1=0=q_2$, the metric becomes AdS$_3\times S^3\times S^3\times \mathbb R$, which is the decoupling limit of the string/fivebrane/fivebrane configuration \cite{Cowdall:1998bu}.  The rotations associated with parameters $(q_1,q_2)$ turn the AdS$_3$ into the G$_\alpha$.  We thus expect that there should be a rotating string/fivebrane/fivebrane configuration whose decoupling limit gives rise to our ten-dimensional solution (\ref{10dsol}).  If we set either $g_1=0$ or $g_2=0$, but not both, the associated $S^3$ is flatten to become $\mathbb R^3$.  The metric configuration becomes $G_\alpha\times S^3\times \mathbb R^4$.  The heterotic string solution of $G_\alpha \times S^3\times K_3$ was first constructed in \cite{Israel:2003cx}.

To study the global structure, we first denote $r_0$ as the largest root of $f(r)$.  Shifting the coordinates as
\be
t\rightarrow t - r_0\phi\,,\qquad \psi_i \rightarrow \psi_i - g_i q_i r_0\phi\,,
\ee
we find that the metric is singular at $r=r_0$, where the degenerate Killing vector is purely spatial $\xi=\partial_\phi$. The absence of a conical singularity requires that
\be
\Delta \phi = \fft{4\pi}{\sqrt{\alpha} r_0^2 f'(r_0)}\,.
\ee
Thus in this system, there are three periodic coordinates $\phi$, and $(\psi_1,\psi_2)$, with $\Delta\psi_1=4\pi=\Delta \psi_2$. The coordinate $t$ on the other hand is not required to be periodic.  The analysis of CTCs in ten dimensions becomes more subtle.  Note that we have
\be
g_{\phi\phi} = \fft{\alpha r_0^2 f'(r_0)}{g_1^2 + g_2^2} (r-r_0)\ge 0\,,
\ee
for the region $r\ge r_0$; however, naked CTCs still exist.  One way to see this is to consider the general periodic Killing vector
\be
\xi=\beta \fft{\partial}{\partial\phi} + \gamma_1 \fft{\partial}{\partial \psi_1} +
\gamma_2 \fft{\partial}{\partial \psi_2}\,.
\ee
The absence of naked CTCs requires that $\xi^2\ge 0$ for all real $(\beta,\gamma_1,\gamma_2)$ in the $r\ge r_0$ region.  This can be easily established not true.  Negative modes arise for large enough $r$.

   A simpler way to see that naked CTCs exist is the follows.  Let $r=r_*$ such that
$r_*^2(\alpha f(r_*)-1)<0$,  which is always achievable since $0<\alpha<1$. Now making a coordinate shifting $\psi_i\rightarrow \psi_i - g_i q_i r_*\phi$, then we have $g_{\phi\phi}(r_*)=r_*^2(\alpha f(r_*)-1)<0$.

\subsection{AdS$_3\times S^3$ bundle over $S_\alpha^3$}

We now consider the case with negative $g_1^2=-\hat g_1^2$ and $\hat g_1^2 - g_2^2>0$.  Performing some appropriate analytical continuation of the coordinates on the solution (\ref{10dsol}) and then dropping the hat symbol, we have
\bea
ds_{10}^2 &=& \fft{1}{g_1^2 - g_2^2} \Big((d\psi + \cos\theta\, d\phi)^2 + \alpha \sin^2\theta\, d\phi^2 + d\theta^2+ dz^2\Big)\nn\\
&&+ \fft{1}{g_1^2} \Big( (d\psi_1 + \rho dt + g_1 q_1 \cos\theta\, d\phi)^2 +\fft{d\rho^2}{\rho^2+1} -(\rho^2+1)\,dt^2\Big) \nn\\
&&+ \fft{1}{g_2^2} \Big( (d\psi_2 + \cos\theta_2\, d\phi_2 + g_2 q_2 \cos\theta\, d\phi)^2 + d\theta_2^2 + \sin^2\theta_2\,d\phi_2^2\Big)\\
F_\3 &=& -\fft{1}{g_1^2 - g_2^2} d\psi\wedge d\cos\theta\wedge d\phi + \fft{1}{g_1^2} d\rho\wedge dt\wedge
(d\psi_1 + \rho\,dt + g_1 q_1 \cos\theta\, d\phi)\nn\\
&& - \fft{\sin\theta_2}{g_2^2} d\theta_2\wedge d\phi_2\wedge
(d\psi_2 + \cos\theta_2\,d\phi_2 + g_2 q_2 \cos\theta\, d\phi)\nn\\
&& -\fft{q_1}{g_1} d\cos\theta\wedge d\phi \wedge (d\psi_1 + \rho dt)
-\fft{q_2}{g_2} d\cos\theta\wedge d\phi \wedge (d\psi_2 + \cos\theta_2\, d\phi_2)\,,
\eea
where $\alpha=1 + (g_1^2-g_2^2)(q_1^2 + q_2^2)$.  The solution describes a direct production of a line of coordinate $z$ and a nine-dimensional metric of an AdS$_3\times S^3$ bundle over squashed 3-sphere $S_\alpha^3$.  Again the configuration involves both electric string and magnetic fivebrane/fivebrane charges
\bea
\hbox{electric}:&& Q_1\sim\fft{g_1}{\sqrt{\alpha}\,g_2^2(g_1^2-g_2^2)^2}\,,\nn\\
\hbox{magnetic}:&& Q_{5/5} \sim \fft{1}{\alpha\, (g_1^2 - g_2^2)} + \fft1{g_2^2}\,.
\eea
Note that $\Delta\psi\sim \Delta\phi\sim 1/\sqrt{\alpha}$.
When $q_1=0=q_2$, the metric is again AdS$_3\times S^3\times S^3$, equivalent to the previous static case.  For non-vanishing $q_i$'s, the brane configuration is not clear and it deserves further study.  Interestingly, the limit of $g_1=g_2$ leads to the well-known AdS$_3\times S^3\times \mathbb R^4$ vacuum of string theory, and the limit $g_2=0$ gives rise to AdS$_3\times S_\alpha^3\times \mathbb R^4$.

\section{Conclusions}

Four-dimensional G\"odel metrics of G$_\alpha\times \mathbb{R}$ are perhaps the simplest solutions that exhibit naked CTCs with no globally spatial-like Cauchy horizon.  In this paper, we showed that the G\"odel metrics could arise as exact solutions in Lagrangian formalism.  We constructed EMA and EPA theories that admit G\"odel solutions.  We also showed that G\"odel universe could emerge from Freedman-Schwarz $SU(2)\times SU(2)$ gauge supergravity.  This allows us to give exact embeddings of the G\"odel metrics in string theories.  The ten-dimensional solution describes a direct product of a line and an $S^3\times S^3$ bundle over G$_\alpha$. Classically, we find that naked CTCs persist in higher dimensions. (In \cite{Israel:2003cx}, string quantization was performed on G$_\alpha\times S^3\times K_3$ and it was demonstrated that CTCs can resolved by the quantum effects.) For some appropriate choice of parameters, the nine-dimensional metric can describe an AdS$_3\times S^3$ bundle over a squashed 3-sphere $S_\alpha^3$, in which case, there is no CTC.  In the suitable limit, the solution becomes the supersymmetric AdS$_3\times S^3\times \mathbb R^4$ vacuum.

The scaling symmetry of the metric (\ref{gvac}) resembles that of the anti-de Sitter spacetimes.  This is suggestive that there may exist a boundary field theory at the $r\rightarrow \infty$ boundary of G\"odel universe. The exact embedding of the G\"odel metrics in string theory, as the decoupling limit of the rotating D1/D5/D5 intersection, provides a tool of investigating the boundary field theory in the context of string theory.

\section*{Acknowlegement}

S.-L.L. and H.W. are supported in part by NSFC under Grants NO. 11575022 and NO. 11175016. X.-H.F. and H.L. are supported in part by NSFC grants NO. 11175269, NO. 11475024 and NO. 11235003.

\end{document}